\documentclass{article}
\usepackage{spconf}

\usepackage{amsmath,amssymb,bbm}

\usepackage[breaklinks=true,letterpaper=true,bookmarks=false,hidelinks]{hyperref}
\usepackage{cleveref}
\usepackage{url}

\usepackage{graphicx}
\usepackage{caption}
\usepackage{subcaption}

\usepackage{algorithm}
\usepackage{algorithmic}

\title{Unsupervised Style and Content Separation by Minimizing Mutual Information for Speech Synthesis}
\name{Ting-Yao Hu$^{1*}$, Ashish Shrivastava$^{2}$, Oncel Tuzel$^{2}$, Chandra Dhir$^{2}$
\thanks{$^*$Work done during summer internship at Apple. Emails: tingyaoh@andrew.cmu.edu, \{ashish.s, otuzel, cdhir\}@apple.com}
\vspace{-0.15in}
}
\address{
$^1$Carnegie Mellon University \;\;\;\;\;\;\;\; $^2$Apple Inc. }
\graphicspath{{./Figures/}}

\begin{document}
%
\maketitle
\begin{abstract}
We present a method to generate speech from input text and a style vector that is extracted from a reference speech signal in an unsupervised manner, i.e., no style annotation, such as speaker information, is required.
Existing unsupervised methods, during training, generate speech by computing style from the corresponding ground truth sample and use a decoder to combine the style vector with the input text.
Training the model in such a way leaks content information into the style vector.
The decoder can use the leaked content and ignore some of the input text to minimize the reconstruction loss.
At inference time, when the reference speech does not match the content input, the output may not contain all of the content of the input text.
We refer to this problem as ``content leakage", which we address by explicitly estimating and minimizing the mutual information between the style and the content through an adversarial training formulation.
We call our method MIST - \textbf{M}utual \textbf{I}nformation based \textbf{S}tyle Conten\textbf{t} Separation.
The main goal of the method is to preserve the input content in the synthesized speech signal, which we measure by the word error rate (WER) and show substantial improvements over state-of-the-art unsupervised speech synthesis methods.
\end{abstract}

\begin{keywords}
Unsupervised style-content separation, mutual information estimation, controllable speech synthesis.
\end{keywords}
\section{Introduction}
Although neural network based text-to-speech (TTS) models \cite{Shen17_tacotron2, arik17a_deepvoice, Li2019NeuralSS} can produce high quality speech, an input text is mapped to only one speech signal.
In reality, one text may correspond to different speech outputs due to variations in the speaker identity, the speaking style, prosody, or the environmental factors.
Researchers have addressed this problem by providing an additional reference speech signal to control the style of the generated speech \cite{style_tokens_wang18h, hsu2018hierarchical,jia2018transfer,Ma2019iclr}.
In these controllable TTS methods, the reference speech is encoded into an embedding (called a style vector) that is input with the content features to a speech decoder.
Most recent works in this direction generate the style vector that uses speaker identity \cite{hsu2018hierarchical,jia2018transfer,Ma2019iclr, vasquez2019_melnet}, which may be hard to extend to the cases where the speaker information is not available (e.g. a new language, a new environment, or due to privacy reasons).
Furthermore, the user information requires additional annotations to use the unlabeled audio data.
To overcome this limitation, Global Style Token (GST) method~\cite{style_tokens_wang18h} learns speaker embeddings in an unsupervised manner by jointly training the style encoder network as well as the encoder-decoder part of the TTS model, while minimizing the reconstruction loss.
Since the whole system is learned in an end-to-end fashion using just the reconstruction loss, some of the content information is leaked into the style vector, which the decoder can use to reconstruct the ground-truth speech features.

Following~\cite{style_tokens_wang18h}, we compute style vectors using a set of trainable vectors called style tokens, which are linearly combined using style coefficients generated from the input reference speech.
Style tokens are trainable parameters that are optimized together with the TTS network parameters.
To compute the style coefficients, we use an additional style encoder that is trained jointly with the TTS model.
The style coefficients are passed through a Softmax layer (so that they sum to $1$) before computing the style vector with them.
Furthermore, for computational efficiency, we use Transformer TTS~\cite{NIPS2017_transformers, Li2019NeuralSS} for the content encoder and decoder.
This model uses self-attention~\cite{NIPS2017_transformers} and does not have any recurrent connections, which is significantly faster to train compared to LSTM-based models such as Tacotron 2~\cite{Shen17_tacotron2}.
During training, text is given as the content input and the corresponding mel-spectrogram is used as reference speech for encoding the style.

In this setup, the desired output is the same as the reference input for style encoding, which causes some of the content information to leak into the style vector.
This leaked content can be used by the decoder to reconstruct the speech while ignoring the actual content input.
At inference time, when the reference speech has different content from the input text, the decoder expects the content from the style vector and ignores some part of the content text.
We refer to this problem as ``content-leakage" which results from having the same style input as the desired output during training.
Ideally, the style vector should not be able to reconstruct the content vector, i.e., there should be no information about the content in the style vector.
To this end, we minimize the mutual information between the style and the content vectors.
We estimate the mutual information between the style and the content vectors using Mutual Information Neural Estimation (MINE) proposed in~\cite{belghazi18a_MINE}.
The MINE algorithm computes a lower bound of the mutual information using a neural network,
which is optimized to maximize this lower bound.
We alternate between maximizing the lower bound (i.e., estimating the mutual information) and minimizing the estimated mutual information and the reconstruction loss.
The maximization problem is solved w.r.t. the MINE network, while the minimization problem is solved w.r.t. the style encoder, the content encoder, and the decoder.

To summarize our contributions, we prevent content leakage for controllable TTS by minimizing the mutual information between the style and the content vectors.
We evaluate our method quantitatively and qualitatively and outperform state-of-the-art unsupervised controllable TTS methods.

\section{Related Works}
Recent neural TTS methods, such as Tacotron 2 \cite{Shen17_tacotron2}, MelNet \cite{vasquez2019_melnet}, Deep Voice 3 \cite{ping2018_deepvoice3}, and TransformerTTS \cite{Li2019NeuralSS}, map input text to speech features (e.g. mel-spectrogram) using a content encoder and a speech decoder.
To recover the original time domain speech signal from the speech features, one can rely on a conventional vocoder such as Griffin Lim algorithm \cite{Griffin84signalestimation}, or a neural network based vocoder, such as WaveNet~\cite{oord2016wavenet} and WaveGlow \cite{prenger2018_waveglow}.
We choose TransformerTTS as our neural TTS backbone because of the substantially reduced training time, and WaveNet~\cite{oord2016wavenet} as our vocoder.

The concept of style and content disentanglement has been explored in many different areas, such as artistic image \cite{gatys2016image}, face attribute manipulation~\cite{lample2017fader}, handwriting~\cite{chen2016infogan}, text generation~\cite{hu2017toward}, and neural TTS~\cite{Ma2019iclr}.
The authors in \cite{Ma2019iclr} follow the idea of obtaining the style information as the gram matrix of feature maps to capture the style in synthesized speech.
Compared to these methods, our approach disentangles the style and the content by explicitly minimizing the mutual information between their latent representations, not the loss of a discriminator.

Neural controllable TTS models \cite{style_tokens_wang18h, hsu2018hierarchical,jia2018transfer,Ma2019iclr} generate speech with the input text content, where the style is given by an input reference speech signal that may not have the same content as the input text.
These models analyze the reference speech signal and extract style information using an additional style encoder, which is parallel to the content encoder of a neural TTS system.
The authors in~\cite{jia2018transfer} incorporate external data to train a discriminative speaker encoder, and transfer the learned encoder to build a multi-speaker TTS system.
The authors in~\cite{hsu2018hierarchical} adopt a variational autoencoder to model both the observed and the latent style attributes.
Global style token (GST) method~\cite{style_tokens_wang18h} maintains a set of style embedding vectors, and constrain a style embedding of reference speech to be a linear combination of this style embedding set.
A recent work \cite{Ma2019iclr} enhances this model by latent attribute reconstruction and GAN training \cite{NIPS2014_GAN}.
Most of the these works require style annotation, such as speaker identity and emotion, in the training stage.
Compared to these methods, our proposed approach is unsupervised, i.e., it does not require style annotations or speaker embeddings.
To the best of our knowledge, the only other neural TTS based unsupervised style and content separation method is by \cite{style_tokens_wang18h}, but this suffers from content leakage.

\section{The Proposed {MIST} Approach}
\begin{figure}
    \centering
    \includegraphics[width=.99\linewidth]{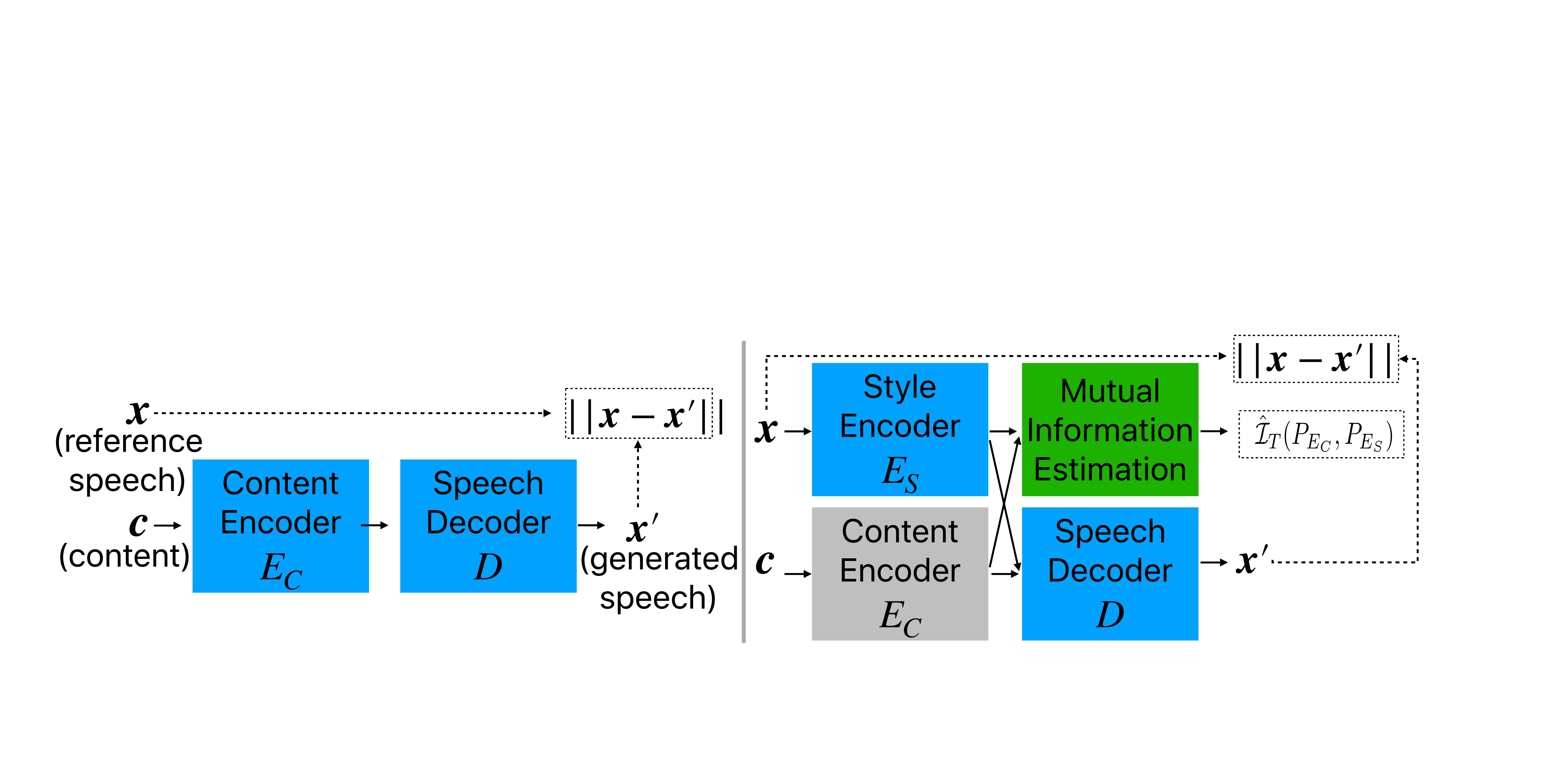} \\
    (a) \qquad \qquad \qquad \qquad \qquad  (b)
    \vspace{-0.1in}
    \caption{Overview of the proposed method. (a) We pre-train the content encoder with a single speaker dataset. (b) With fixed content encoder, we alternate between estimating the mutual information $\hat{\mathcal I}$ (green box) and optimizing style encoder + speech decoder (blue boxes) to minimize the reconstruction loss, $\|\boldsymbol x - \boldsymbol x'\|$, and the $\hat{\mathcal I_T}$.}
    \label{fig:method}
    \vspace{-0.1in}
\end{figure}

The proposed method, shown in Figure~\ref{fig:method}, is based on a controllable TTS architecture.
We use a backbone TTS model to pre-train the content encoder, $E_C$ (Figure~\ref{fig:method}(a)).
To this backbone TTS model, we add a style encoder, $E_S$, to extract style vector from the reference speech, and the MI estimator to measure the mutual information between the style and the content vectors (Figure~\ref{fig:method}(b)).
\vspace{-0.1in}

\subsection{Content Encoder Pre-training}
\vspace{-0.05in}
The first step of MIST is content encoder pre-training, which can be simply treated as a neural TTS training process.
It is important to use a single-style dataset in the pre-training process because a multi-style dataset usually has same content spoken in different style (e.g. by different speakers).
Given a set of speech and content pairs, $\{(\boldsymbol x, \boldsymbol c)\}$, we jointly train the content encoder, $E_C$, and speech decoder, $D$, by minimizing the reconstruction loss,
$min_{E_C,D} ||D(E_C(\boldsymbol c))-\boldsymbol x||_1$,
where $\|.\|_1$ is the $\ell_1$ norm. The trained $E_C$ with frozen weights is used in the second stage of our method, while $D$ is re-initialized with random weights.
\vspace{-0.1in}


\subsection{Style and content disentanglement}
\vspace{-0.05in}
In the second step of our method, we train a speech synthesis model that is capable of disentangling the style from the reference speech and generating speech in this style with the content of the input text.
During training, the input content is the same as the content of the reference speech.
Using only the reconstruction loss to update $E_S, E_C$, and $D$, the model suffers from content leakage because the content information in the output can also be extracted from the reference speech.
We disentangle the style and content by minimizing the mutual information (MI) between their hidden representations ($E_S(\boldsymbol x)$ and $E_C(\boldsymbol c)$), so that the style does not contain information about the content.
However, it is not obvious how to compute and minimize the mutual information between two continuous random vectors.
First, we briefly describe a recently proposed method to estimate the mutual information, then we present our novel application to minimize it jointly with the reconstruction loss.

\noindent \textbf{Mutual information neural estimation (MINE)\cite{belghazi18a_MINE}:}
The mutual information, $\mathcal I(\boldsymbol Y, \boldsymbol Z)$, of random variables $\boldsymbol Y$ and $\boldsymbol Z$ is equivalent to the Kullback–Leibler (KL) divergence \cite{Kullback51klDivergence} between their joint distribution, $P_{\boldsymbol Y, \boldsymbol Z}$, and product of marginals, $P_{\boldsymbol Y}*P_{\boldsymbol Z}$, i.e.,
$\mathcal I(\boldsymbol Y,\boldsymbol Z) = D_{KL}(P_{\boldsymbol Y, \boldsymbol Z} || P_{\boldsymbol Y}*P_{\boldsymbol Z}).$
Using this fact, MINE\cite{belghazi18a_MINE} method constructs a lower bound of mutual information based on Donsker-Varadhan representation of KL divergence \cite{donsker1983asymptotic}:
\vspace{-0.1in}
\begin{equation*}
  \vspace{-0.05in}
\mathcal I(\boldsymbol Y,\boldsymbol Z) \geq \hat{\mathcal I}_T(\boldsymbol Y,\boldsymbol Z) = \sup_T E_{P_{\boldsymbol Y, \boldsymbol Z}}[T]-\log (E_{P_{\boldsymbol Y}*P_{\boldsymbol Z}}[e^T]),
\label{eq:mine_obj}
\end{equation*}
where $T$ can be any function that makes the two expectations in the above equation finite.
The authors in \cite{belghazi18a_MINE} propose to use a deep neural network for $T$,
which allows us to estimate the mutual information between $\boldsymbol Y$ and $\boldsymbol Z$ by maximizing this lower bound with respect to $T$ through gradient descent.

\vspace{0.05in}
\noindent \textbf{Style and content separation with MI minimization:}
We minimize the the reconstruction loss along with the estimated mutual information between the style and the content vectors.
Since the MI is always non-negative, we clip the estimated mutual information to zero if it is negative.
The clipped value is not only a better estimate of the mutual information than the non-clipped one (because the true MI is always non-negative), it also avoids minimizing a function that is unbounded from below.
In one experiment, we found that by clipping the performance of the speech recognition on the generated data can be improved by approximately $30\%$.
Thus, the overall objective function is a min-max problem where we maximize the lower-bound of MI, $\hat{\mathcal I}$, w.r.t. $T$ and minimize the MI and the reconstruction loss w.r.t. $D$ and $E_S$,
\vspace{-0.05in}
\begin{align}
\min_{E_S,D}\max_T & \big \{||D(E_C(\boldsymbol c), E_S(\boldsymbol x)) - \boldsymbol x||_1 \nonumber \\
&+ \lambda * \max(0, \hat{\mathcal I}_T(E_C(\boldsymbol c), E_S(\boldsymbol x))) \big \},
\label{eq:opt_prob}
\end{align}
where $\lambda$ is a hyper-parameter that balances the two losses.
In our experiments, we set $\lambda=0.1$ and found the algorithm to be insensitive to different values of $\lambda$, as shown later in Section~\ref{sec:lambda_sensitive_analysis}.
Similar to common GAN training, we update the speech synthesis model ($E_S, D$) and the MI estimator function, $T$, alternatively in each step of the training.
Since $E_C(\boldsymbol c_i)$ is a sequence of vectors of varying length, we randomly sample one of the content vectors to compute the mutual information.
By optimizing \eqref{eq:opt_prob}, we can jointly ensure the quality of speech feature reconstruction, and make the information extracted from $E_C$ and $E_S$ independent to each other.
We summarize the training method in Algorithm~\ref{alg:algo_steps}.

\begin{algorithm}
\caption{Pseudocode for the proposed MIST training}
\textbf{Input:} Pairs of speech and text ($\boldsymbol x_i$, $\boldsymbol c_i$).\\
\textbf{Output:} $E_C, D, E_S.$

\begin{algorithmic}[1]
\STATE $E_C, D \gets \arg\min_{E_C,D} \sum_i||D(E_C(\boldsymbol c_i))-\boldsymbol x_i||_1$ 
\STATE $E_S, D, T \gets$ initialization with random weights
\WHILE {$E_S, D, T$ not converged}
\STATE Sample a mini-batch of $(\boldsymbol x_i, \boldsymbol c_i), i=1,2...,b$.
\STATE $\{\boldsymbol y_i\} \gets \{E_C(\boldsymbol c_i)| i=1,2,...,b\}$
\STATE $\{\hat{\boldsymbol y}_i\} = $ random permutation of $\{\boldsymbol y_i\}$
\STATE $\{\boldsymbol z_i\} \gets \{E_S(\boldsymbol x_i) | i=1,2,...,b\}$
\STATE $\mathcal L_{MI} = \frac{1}{b}\sum_{i=1}^b T(\boldsymbol y_i, \boldsymbol z_i) - \log (\frac{1}{b}\sum_{i=1}^b e^{T(\hat{\boldsymbol y}_i, \boldsymbol z_i)})$
\STATE $\mathcal L = \frac{1}{b}\sum_{i=1}^b ||D(\boldsymbol y_i, \boldsymbol z_i)- \boldsymbol x_i||_1 + \lambda* \max(0, \mathcal L_{MI}) $
\STATE $D = D - \epsilon \nabla_D \mathcal L$
\STATE $E_S = E_S - \epsilon \nabla_{E_S} \mathcal L$
\STATE $T = T + \epsilon \nabla_T \mathcal L_{MI}$
\ENDWHILE
\end{algorithmic}
\label{alg:algo_steps}
\end{algorithm}

The pre-training for the content encoder is also a crucial step for style and content disentanglement.
If the content encoder is not pre-trained, the model could learn to capture part of the content from style encoder, and still minimize the mutual information between $E_C(\boldsymbol c)$ and $E_S(\boldsymbol x)$.




%
\section{Experiments}
To evaluate the effectiveness of MIST on preventing content leakage and the quality of the generated speech, we conduct qualitative and quantitative studies on the VCTK \cite{vctk_2012} and the LibriTTS \cite{zen2019libritts} datasets.
The VCTK dataset contains $44$ hours of clean speech from $109$ speakers, and
LibriTTS \cite{zen2019libritts} is a large-scale corpus with $585$ hours of English speech, which are recorded from 2,456 speakers.
For LibriTTS, we use the train-clean-360 set to learn our model.
We also use LJSpeech dataset \cite{ljspeech17}, which consists of 13,100 short audio clips from a single speaker, for pre-training the content encoder.
For fair comparison, our implementations of the baseline methods also use this pre-trained content encoder.

\noindent \textbf{Baselines:}
We compare our method with the unsupervised method by \cite{style_tokens_wang18h} that proposed to use global style tokens (GST).
The original GST method uses an LSTM based Tacotron 2~\cite{Shen17_tacotron2} as the TTS backbone and an LSTM encoder for computing the style coefficients.
For training efficiency and fair comparison, in our implementation of GST, we use Transformer TTS~\cite{Li2019NeuralSS} for the content encoder and the decoder, and replace the LSTM with max-pooling for computing the style coefficients.
We refer to our implementation of this method as GST\textsuperscript*.
We also compare our method with a recently proposed supervised controllable speech synthesis method~\cite{Ma2019iclr}.
This method uses speaker identities for optimizing the style vectors.
Same as for GST method, our implementation of this method uses Transformer TTS for the TTS backbone.
We refer to our implementation of this method as ~\cite{Ma2019iclr}\textsuperscript*.
All the baseline methods use pre-trained content encoder.

\subsection{Quantitative study}
Since the main objective of MIST algorithm is to improve the content quality of the generated speech,
we objectively evaluate the performance by measuring the content quality using an  ASR (automatic speech recognition) algorithm.
Following ~\cite{Ma2019iclr}, we adopt WaveNet \cite{oord2016wavenet} as the acoustic model in the ASR, and compute word error rate (WER), as a metric for content preservation ability of the model.
The Wavenet model is trained on real speech data with Connectionist Temporal Classification (CTC) loss \cite{graves2006connectionist} between the predicted and the ground truth characters.
For the VCTK dataset, this model achieves a WER of $0.08$ on the held-out real data.
In the testing phase, we prepare $100$ pairs of unmatched text content and reference speech $(\boldsymbol c, \boldsymbol x)$ for both datasets, and report the performance of ASR as WER.
A smaller WER indicates less content leakage.
We present our results in Table~\ref{tb:result_asr_sr}, where the proposed method improves the WER compared to state-of-the-art methods.


\begin{table}[]
  \centering
\begin{tabular}{c|c|c|c}
                & VCTK        &  LibriTTS  & S / U    \\ \hline
\cite{Ma2019iclr}\textsuperscript*  & $34.6\pm0.9$     & $40.0$  & S \\
GST\textsuperscript* ($50$ tokens)        & $50.3\pm4.2$      & $47.7\pm1.2$  & U     \\
GST\textsuperscript* ($10$ tokens)        & $35.7\pm0.5$      & $40.3\pm1.7$ & U      \\
MIST ($50$ tokens) & $29.3\pm1.7$      & $44.3\pm1.7$    & U      \\
MIST ($10$ tokens) & $\mathbf{20.3\pm1.2}$       & $\mathbf {33.3\pm1.2}$ & U \\ \hline
\end{tabular}
\vspace{-0.1in}
\caption{
Word error rate (WER) on the synthesized speech for the VCTK and the LibriTTS datasets.
As shown by the smaller WER, the proposed MIST algorithm preserves the content better than the baselines.
The last column shows whether the method is supervised (S) or unsupervised (U).
}
\label{tb:result_asr_sr}
\end{table}

\noindent \textbf{Sensitivity Analysis of the Hyper-parameter $\lambda$: }
\label{sec:lambda_sensitive_analysis}
To investigate the sensitivity of the hyper-parameter $\lambda$, the combination weight between reconstruction loss and MI minimization, we evaluate our model with different values of $\lambda$.
In this set of experiments, we use $10$ tokens in the style encoder, and measure the WER with the VCTK dataset.
For a range of $\lambda$ values, $0.05-0.5$, the WER was $0.20-0.22$, which shows that MIST is insensitive to exact value of this hyper-parameter.

 \begin{figure}[ht]
      \centering
      \vspace{-0.2in}
      \includegraphics[width=.8\linewidth]{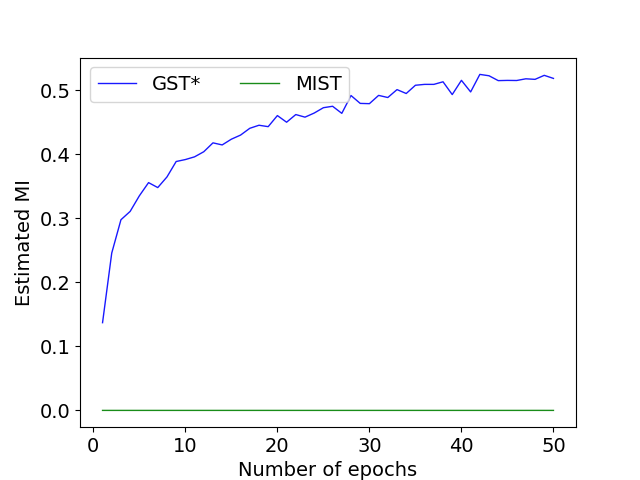}
      \vspace{-0.1in}
      \caption{
        The MI estimates, for frozen TTS models, shown as a function of the training epochs of the MINE.
        }
      \label{fig:mi_vis_post}
  \end{figure}

\noindent \textbf{Analysis of the mutual information loss: }
After training the speech synthesis model, we expect the mutual information between the style vectors, ($E_S(\boldsymbol x)$), and the content vectors, ($E_C(\boldsymbol c)$), be small.
To verify this hypothesis, we estimate the mutual information between the two random variables (i.e. the style vectors and the content vectors) from our trained model (with frozen weights) using the MINE algorithm, which is shown in Figure~\ref{fig:mi_vis_post} as function of training epochs.
The MINE algorithm optimizes the MINE neural network, $T$, according to Equation~\eqref{eq:mine_obj} and keeps all other parts ($D, E_S, E_C$) fixed.
The MI estimate stays close to $0$ for more than $50$ epoch with our model, while it increases immediately with the GST\textsuperscript* model.

 \subsection{Qualitative Study}

To evaluate the quality of the synthesized speech, we conducted a user study with $6$ subjects performing a total of $150$ tests.
 Each test consists of a reference speech, a text content (not matching with the content of the reference speech), and two synthesized speech samples from GST\textsuperscript* and MIST, respectively.
The order of both speech samples were randomized for each test.
The participants of the study were asked two questions: (1)~which synthesized speech preserves content better, and (2)~which is more similar to reference speech in terms of style.
There were three choices for each question: (1)~synthesized speech 1 is better, (2)~synthesized speech 2 is better, and (3)~both outputs are the same.
 The results of their ratings are illustrated in Table~\ref{tb:qualitative_user_test}.
From these results, we can see that MIST preserves content of the input text better, as supported by the better ASR results in Table~\ref{tb:result_asr_sr}, and also preserves the style of the reference speech better, compared to the baseline, GST\textsuperscript*, method.

 \begin{table}[]
   \centering
 \begin{tabular}{c|c|c}
                 & Content          &  Style       \\
                 &  Preservation         &   Preservation      \\ \hline
 Both methods are same  & $29.3$      & $41.3$        \\
 Baseline (GST\textsuperscript*) is better      & $26.0$      & $17.3$         \\
 MIST is better       & $\mathbf{44.7}$      & $\mathbf{41.3}$       \\   \hline
 \end{tabular}
 \vspace{-0.1in}
 \caption{Qualitative evaluation:
 The numbers in the first row indicate percentage of time both the methods are rated the same.
 The second and third row are the percentage of time the method in first column is rated better.
}
 \label{tb:qualitative_user_test}
 \end{table}


\section{Conclusion}
We proposed an unsupervised mutual information minimization based content and style separation for speech synthesis.
In each training step, we estimated the mutual information between the style and the content, and minimized it along with the reconstruction loss.
We showed that such training strategy reduces content leakage and results in substantially better WER compared to the baseline approaches.


\bibliographystyle{IEEEbib}
\small
\bibliography{mybib}

\begin{thebibliography}{10}

\bibitem{Shen17_tacotron2}
J.~{Shen}, R.~{Pang}, R.~J. {Weiss}, M.~{Schuster}, N.~{Jaitly}, Z.~{Yang},
  Z.~{Chen}, Y.~{Zhang}, Y.~{Wang}, R.~{Skerrv-Ryan}, R.~A. {Saurous},
  Y.~{Agiomvrgiannakis}, and Y.~{Wu},
\newblock ``Natural {TTS} synthesis by conditioning wavenet on mel spectrogram
  predictions,''
\newblock in {\em Proc. ICASSP}, 2018.

\bibitem{arik17a_deepvoice}
Sercan~{\"O}. Ar{\i}k, Mike Chrzanowski, Adam Coates, Gregory Diamos, Andrew
  Gibiansky, Yongguo Kang, Xian Li, John Miller, Andrew Ng, Jonathan Raiman,
  Shubho Sengupta, and Mohammad Shoeybi,
\newblock ``Deep voice: Real-time neural text-to-speech,''
\newblock in {\em Proc. ICML}, 2017.

\bibitem{Li2019NeuralSS}
Naihan Li, Shujie Liu, Yanqing Liu, Sheng Zhao, and Peng Shi,
\newblock ``Neural speech synthesis with transformer network,''
\newblock in {\em AAAI}, 2019.

\bibitem{style_tokens_wang18h}
Yuxuan Wang, Daisy Stanton, Yu~Zhang, RJ-Skerry Ryan, Eric Battenberg, Joel
  Shor, Ying Xiao, Ye~Jia, Fei Ren, and Rif~A. Saurous,
\newblock ``Style tokens: Unsupervised style modeling, control and transfer in
  end-to-end speech synthesis,''
\newblock in {\em Proc. ICML}, 2018.

\bibitem{hsu2018hierarchical}
Wei-Ning Hsu, Yu~Zhang, Ron~J Weiss, Heiga Zen, Yonghui Wu, Yuxuan Wang, Yuan
  Cao, Ye~Jia, Zhifeng Chen, Jonathan Shen, et~al.,
\newblock ``Hierarchical generative modeling for controllable speech
  synthesis,''
\newblock {\em Proc. ICLR}, 2019.

\bibitem{jia2018transfer}
Ye~Jia, Yu~Zhang, Ron Weiss, Quan Wang, Jonathan Shen, Fei Ren, Patrick Nguyen,
  Ruoming Pang, Ignacio~Lopez Moreno, Yonghui Wu, et~al.,
\newblock ``Transfer learning from speaker verification to multispeaker
  text-to-speech synthesis,''
\newblock in {\em Proc. NIPS}, 2018.

\bibitem{Ma2019iclr}
Shuang Ma, Daniel Mcduff, and Yale Song,
\newblock ``A generative adversarial network for style modeling in a
  text-to-speech system,''
\newblock in {\em Proc. ICLR}, 2019.

\bibitem{vasquez2019_melnet}
Sean Vasquez and Mike Lewis,
\newblock ``Melnet: A generative model for audio in the frequency domain,''
\newblock {\em arXiv preprint arXiv:1906.01083}, 2019.

\bibitem{NIPS2017_transformers}
Ashish Vaswani, Noam Shazeer, Niki Parmar, Jakob Uszkoreit, Llion Jones,
  Aidan~N Gomez, \L~ukasz Kaiser, and Illia Polosukhin,
\newblock ``Attention is all you need,''
\newblock in {\em Proc. NIPS}. 2017.

\bibitem{belghazi18a_MINE}
Mohamed~Ishmael Belghazi, Aristide Baratin, Sai Rajeshwar, Sherjil Ozair,
  Yoshua Bengio, Aaron Courville, and Devon Hjelm,
\newblock ``Mutual information neural estimation,''
\newblock in {\em Proc. ICML}, 2018.

\bibitem{ping2018_deepvoice3}
Wei Ping, Kainan Peng, Andrew Gibiansky, Sercan~O. Arik, Ajay Kannan, Sharan
  Narang, Jonathan Raiman, and John Miller,
\newblock ``Deep voice 3: 2000-speaker neural text-to-speech,''
\newblock in {\em Proc. ICLR}, 2018.

\bibitem{Griffin84signalestimation}
Daniel~W. Griffin, Jae, S.~Lim, and Senior Member,
\newblock ``Signal estimation from modified short-time fourier transform,''
\newblock {\em IEEE Trans. Acoustics, Speech and Sig. Proc}, 1984.

\bibitem{oord2016wavenet}
Aaron van~den Oord, Sander Dieleman, Heiga Zen, Karen Simonyan, Oriol Vinyals,
  Alex Graves, Nal Kalchbrenner, Andrew Senior, and Koray Kavukcuoglu,
\newblock ``Wavenet: A generative model for raw audio,''
\newblock {\em arXiv preprint arXiv:1609.03499}, 2016.

\bibitem{prenger2018_waveglow}
Ryan Prenger, Rafael Valle, and Bryan Catanzaro,
\newblock ``{WaveGlow}: A flow-based generative network for speech synthesis,''
\newblock {\em arXiv preprint arXiv:1811.00002}, 2018.

\bibitem{gatys2016image}
Leon~A Gatys, Alexander~S Ecker, and Matthias Bethge,
\newblock ``Image style transfer using convolutional neural networks,''
\newblock in {\em Proc. CVPR}, 2016.

\bibitem{lample2017fader}
Guillaume Lample, Neil Zeghidour, Nicolas Usunier, Antoine Bordes, Ludovic
  Denoyer, et~al.,
\newblock ``Fader networks: Manipulating images by sliding attributes,''
\newblock in {\em Proc. NIPS}, 2017.

\bibitem{chen2016infogan}
Xi~Chen, Yan Duan, Rein Houthooft, John Schulman, Ilya Sutskever, and Pieter
  Abbeel,
\newblock ``{InfoGAN}: Interpretable representation learning by information
  maximizing generative adversarial nets,''
\newblock in {\em Proc. NIPS}, 2016.

\bibitem{hu2017toward}
Zhiting Hu, Zichao Yang, Xiaodan Liang, Ruslan Salakhutdinov, and Eric~P Xing,
\newblock ``Toward controlled generation of text,''
\newblock in {\em Proc. ICML}, 2017.

\bibitem{NIPS2014_GAN}
Ian Goodfellow, Jean Pouget-Abadie, Mehdi Mirza, Bing Xu, David Warde-Farley,
  Sherjil Ozair, Aaron Courville, and Yoshua Bengio,
\newblock ``Generative adversarial nets,''
\newblock in {\em Proc. NIPS}. 2014.

\bibitem{Kullback51klDivergence}
S.~Kullback and R.~A. Leibler,
\newblock ``On information and sufficiency,''
\newblock {\em Ann. Math. Statist.}, 1951.

\bibitem{donsker1983asymptotic}
Monroe~D Donsker and SR~Srinivasa Varadhan,
\newblock ``Asymptotic evaluation of certain markov process expectations for
  large time. iv,''
\newblock {\em Communications on Pure and Applied Mathematics}.

\bibitem{vctk_2012}
Junichi Yamagishi,
\newblock ``English multi-speaker corpus for cstr voice cloning toolkit,,''
\newblock 2012.

\bibitem{zen2019libritts}
Heiga Zen, Viet Dang, Rob Clark, Yu~Zhang, Ron~J Weiss, Ye~Jia, Zhifeng Chen,
  and Yonghui Wu,
\newblock ``{LibriTTS}: A corpus derived from librispeech for text-to-speech,''
\newblock {\em arXiv preprint arXiv:1904.02882}, 2019.

\bibitem{ljspeech17}
Keith Ito,
\newblock ``The lj speech dataset,''
  \url{https://keithito.com/LJ-Speech-Dataset/}, 2017.

\bibitem{graves2006connectionist}
Alex Graves, Santiago Fern{\'a}ndez, Faustino Gomez, and J{\"u}rgen
  Schmidhuber,
\newblock ``Connectionist temporal classification: labelling unsegmented
  sequence data with recurrent neural networks,''
\newblock in {\em Proc. ICML}, 2006.

\end{thebibliography}

\end{document}